\documentclass[
preprintnumbers,
preprint,
a4paper,
showpacs, 
amsmath, 
amssymb, 
floatfix,
nofootinbib,
]{revtex4}

\usepackage[dvipdfmx]{graphicx}
\begin{document}

%<<<<<<<<<<<<< TITLE >>>>>>>>>>>>>>>%
\title{Time evolution of a thin black ring via Hawking radiation}

%<<<<<<<<<<<<< AUTHOR  >>>>>>>>>>>>>>>%
\author{Mitsuhiro Matsumoto$^{(1)}$}
\author{Hirotaka Yoshino$^{(2)}$}
\author{Hideo Kodama$^{(1,2)}$}

%<<<<<<<<<<<<< AFFILIATION >>>>>>>>>>>>>>>%

\affiliation{$^{(1)}$Department of Particle and Nuclear Physics, 
The Graduate University for Advanced Studies (SOKENDAI), \\ Tsukuba 305-0801, Japan}

\affiliation{$^{(2)}$Theory Center, 
Institute of Particles and Nuclear Studies, 
KEK, Tsukuba, Ibaraki, 305-0801, Japan}

%<<<<<<<<<<<<< Preprint NO >>>>>>>>>>>>>>>%
\preprint{KEK-TH-1550}

%<<<<<<<<<<<<< DATE >>>>>>>>>>>>>>>%
\date{December 3, 2013}

%<<<<<<<<<<<<< pacs NO >>>>>>>>>>>>>>>%

%
%======================================%
%<<<<<<<<<<<<< ABSTRACT >>>>>>>>>>>>>>>%
%======================================%
%
\begin{abstract}

Black objects lose their mass and angular momenta 
through evaporation by Hawking radiation,
and the investigation of their time evolution has a long history.  
In this paper, we study this problem for a five-dimensional doubly spinning black ring. 
The black ring is assumed to emit only massless scalar particles. 
We consider a thin black ring 
with a small thickness parameter, $\lambda\ll 1$, 
which can be approximated by a boosted Kerr string locally. 
We show that a thin black ring evaporates with 
fixing its thickness parameter $\lambda$.  
Further, in the case of an Emparan-Reall black ring, we derive analytic
formulas for the time evolution, which has one parameter to be evaluated numerically. 
We find that the lifetime of a thin black ring is shorter by a factor of $O(\lambda^2)$
compared to a five-dimensional Schwarzschild black hole with the same initial mass.
We also study detailed properties of the Hawking radiation from the thin black ring, 
including the energy and angular spectra of emitted particles. 

\end{abstract}

\maketitle

%
%======================================%
%<<<<<<<<<<<< SECTION I  >>>>>>>>>>>>>>%
%======================================%
%

\section{Introduction}

In four spacetime dimensions, a stationary, asymptotically flat, 
vacuum black hole is completely characterized 
by its mass and spin angular momentum\cite{Carter:1971zc}. 
In particular, the topology of its event horizon must be a sphere \cite{Hawking:1971vc}.
By contrast, in five dimensions,
in addition to the Myers-Perry black hole \cite{Myers:1986un} 
which is a natural generalization of the four-dimensional Kerr black hole,
various exact solutions of black objects with nonspherical horizon topologies
have been found (see \cite{Emparan:2008eg} for a review). 
In this paper, we focus attention to black ring solutions
with the $S^1\times S^2$ horizon topology. 
A black ring solution rotating in the direction of $S^1$ was found 
by Emparan and Reall \cite{Emparan:2001wn}. 
Since a five-dimensional spacetime can have two angular momenta, 
Pomeransky and Sen'kov \cite{Pomeransky:2006bd} extended it to a solution 
with two independent rotation parameters 
(i.e., spinning both in the directions of $S^1$ and $S^2$).

A black hole is known to evaporate due to quantum effects of fields 
in curved spacetime as shown by Hawking \cite{Hawking:1974sw}. 
The rate of mass and angular momentum loss by the Hawking radiation for a Kerr black hole 
was first studied by Page \cite{Page:1976df,Page:1976ki}
taking account of fields with spins 1/2, 1, and 2,
and it was shown that a Kerr black hole spins down 
to a nonrotating black hole regardless of its initial state.
However, Chambers {\it et al.} \cite{Chambers:1997ai} 
(see also \cite{Taylor:1998dk}) 
showed that if only a massless scalar field is taken into account
(i.e., in the absence of fields with nonzero spin),
a four-dimensional Kerr black hole 
evolves to a state with 
the nonvanishing nondimensional rotation parameter, $a/M\simeq 0.555$.
This analysis was extended to five-dimensional Myers-Perry black holes 
by Nomura {\it et al} \cite{Nomura:2005mw}. 
They showed that any such black hole with nonzero rotation parameters
$a$ and $b$ 
evolves toward an asymptotic state with 
$a/M^{1/2}=b/M^{1/2}\simeq0.1975(8/3\pi)^{1/2}$. 
Here, this value is independent of the initial values of $a$ and $b$.

It is interesting to extend these studies to the case of a black ring.
Although the Hawking radiation of black rings has been studied
in various context \cite{Miyamoto:2007,Chen:2007,Zhao:2006,Jiang:2008}, 
the time evolution of a black ring has not been studied up to now. 
The difficulty in this study is that the method of mode 
decomposition of the Klein-Gordon
field in this spacetime is not known
since separation of variables has not been realized,
and therefore, two-dimensional numerical calculations of eigenfunctions
are required. 
In order to avoid this difficulty, we consider 
a thin black ring with a small thickness parameter, $\lambda\ll 1$. 
Here, ``thin'' or the small thickness parameter $\lambda$
means that the $S^2$ radius is much smaller compared to 
the $S^1$ radius.  
In such a situation, a black ring can be approximated
by a boosted black string. 
Then, the separation of variables 
for the scalar field can be done, and  
we have well defined modes. 

Using this thin-limit approximation, 
we give a formulation to study the evolution 
of a thin Pomeransky-Sen'kov black ring by the Hawking radiation, and 
discuss general features that do not depend on details of the greybody factor. 
Then, we apply our method to a special case of the Emparan-Reall black ring
without $S^2$ rotation, and derive a semi-analytic formula 
for the time evolution of the evaporation. 
Here, the formula is semi-analytic in the sense that 
the evolution is expressed by analytic formulas 
but they include one parameter related to the greybody factors 
that have to be evaluated numerically. 
By developing a numerical code, we also determine the value of this parameter
with sufficient numerical accuracy. 

In addition to the time evolution,
we present numerical results on detailed properties 
of the evaporation of a thin Emparan-Reall black ring.
Specifically, we examine the energy and angular spectra 
of emitted particles in the evaporation. 
In order to clarify the property of the energy spectrum
that is specific to the evaporation of a black ring, 
we discuss the results by comparing it with that 
of a four-dimensional Schwarzschild black hole.

This paper is organized as follows. 
In Sec.~II, the black ring solution is reviewed 
and its boosted Kerr string limit is shown. 
In Sec.~III, we derive the equations that determine the emission rates of mass 
and angular momenta of a black ring via Hawking radiation. 
In Sec.~IV, the time evolution of an evaporating black ring is studied. 
In Sec.~V, we present 
the energy and angular spectra of emitted particles
in the evaporation of a thin Emparan-Reall black ring. 
Sec.~VI is devoted to a summary. 
In Appendix~A, we check the validity of 
our numerical result by studying the DeWitt approximation,
where the greybody factor is approximately evaluated 
using the capture condition of null geodesics
in a black string spacetime. 
To simplify the notation, we use the natural units 
$\hbar = c = G = k_B = 1$, where $G$ is the five-dimensional gravitational constant. 

%
%======================================%
%<<<<<<<<<<<< SECTION II  >>>>>>>>>>>>>>%
%======================================%
%
\section{Black Ring}
In this section, we review basic properties of a black ring 
and show its boosted Kerr string limit.
This limit was discussed in the more general case 
of an unbalanced Pomeransky-Sen'kov black ring 
in Ref.~\cite{Chen:2011jb}.

\subsection{Pomeransky-Sen'kov solution}
\begin{figure}[tb]
\centering
{
\includegraphics[width=0.6\textwidth]{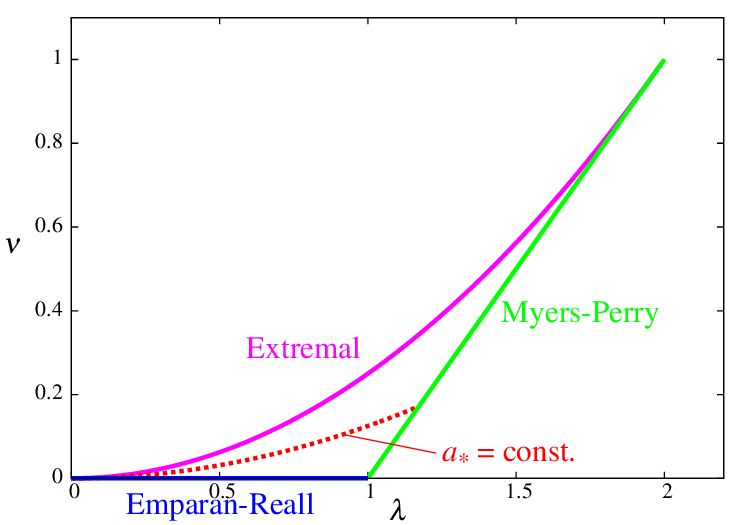}
}
\caption{
The parameter space $(\lambda,\nu)$ of the Pomeransky-Sen'kov solution.
$\lambda$ and $\nu$ can take values in the region surrounded by
solid lines.
The line $\nu=\lambda-1$ is the Myers-Perry black hole limit, 
the line $\nu=0$ is the Emparan-Reall black ring limit, 
and the line $\nu=\lambda^2/4$ is the extremal limit.
The point $\lambda=\nu=0$ corresponds to the boosted Kerr string limit.
The broken line $\nu=a_\ast^2\lambda^2/4$ is the path 
to the point $\lambda=\nu=0$ with a fixed $a_\ast$.
}
\label{Fig:parameter}
\end{figure}

The metric of the Pomeransky-Sen'kov black ring is \cite{Pomeransky:2006bd}
\begin{equation}
\label{PS}
\begin{split}
ds^2=&-\frac{H(y,x)}{H(x,y)}\left(dt+\Omega\right)^2
      -\frac{F(x,y)}{H(y,x)}d\psi^2 - 2\frac{J(x,y)}{H(y,x)}d\psi d\phi + \frac{F(y,x)}{H(y,x)}d\phi^2
   \\&+\frac{2R^2H(x,y)}{\left(x-y\right)^2\left(1-\nu\right)^2}\left(\frac{dx^2}{G(x)}-\frac{dy^2}{G(y)}\right),
\end{split}
\end{equation}
where the 1-form $\Omega$ is
\begin{equation}
\begin{split}
 \Omega=&-\frac{2R\lambda\sqrt{\left(1+\nu\right)^2-\lambda^2}}{H(y,x)}
       \Big[\left(1-x^2\right)y\sqrt{\nu}d\phi
            \\& +\frac{1+y}{1-\lambda+\nu}\left\{1+\lambda-\nu+x^2y\nu\left(1-\lambda-\nu\right)+2\nu x\left(1-y\right)\right\}d\psi
       \Big],
\end{split}
\end{equation}
and the functions $G,H,J$ and $F$ are 
\begin{equation}
G(x)=\left(1-x^2\right)\left(1+\lambda x+\nu x^2\right),
\end{equation}
\begin{equation}
H(x,y)=1+\lambda^2-\nu^2+2\lambda\nu\left(1-x^2\right)y+2x\lambda\left(1-y^2\nu^2\right)+x^2y^2\nu\left(1-\lambda^2-\nu^2\right),
\end{equation}
\begin{equation}
J(x,y)=\frac{2R^2\left(1-x^2\right)\left(1-y^2\right)\lambda\sqrt{\nu}}{\left(x-y\right)\left(1-\nu\right)^2}
       \left[1+\lambda^2-\nu^2+2\left(x+y\right)\lambda\nu-xy\nu\left(1-\lambda^2-\nu^2\right)\right],
\end{equation}
\begin{equation}
\begin{split}
F(x,y)=&\frac{2R^2}{\left(x-y\right)\left(1-\nu\right)^2}
\\&\times\biggl[G(x)\left(1-y^2\right)\left[\left\{(1-\nu)^2-\lambda^2\right\}(1+\nu)+y\lambda\left(1-\lambda^2+2\nu-3\nu^2\right)\right]
\\&\begin{split}
\hspace{8mm} +G(y)\bigl[&2\lambda^2+x\lambda\left\{(1-\nu)^2+\lambda^2\right\}+x^2\left\{(1-\nu)^2-\lambda^2\right\}(1+\nu)
\\&+x^3\lambda\left(1-\lambda^2-3\nu^2+2\nu^3\right)-x^4(1-\nu)\nu\left(-1+\lambda^2+\nu^2\right)\bigl]\biggl].
\end{split}
\end{split}
\end{equation}
Here, we follow the notation of Ref.~\cite{Pomeransky:2006bd}  
except that we choose the signature $(-,+,+,+,+)$ for the metric, 
exchange $\phi$ and $\psi$, and use $R$ instead of $k$.
The coordinate ranges are $-\infty<t<+\infty$,  $0<\phi,\psi<2\pi$,  
$-1\leq x \leq 1$ and $-\infty<y<-1$.
$R$ is a parameter of dimension of length, which determines 
the characteristic scale of the $S^1$ radius.
$\lambda$ and $\nu$ are dimensionless parameters satisfying
$0\leq\nu<1$ and $2\sqrt{\nu}\leq\lambda<1+\nu$,
which determine two nondimensional rotation parameters.
The regular event horizon exists at $y = y_h$, where
\begin{equation}
y_h=\frac{-\lambda+\sqrt{\lambda^2-4\nu}}{2\nu}.
\end{equation}
The solution is asymptotically flat and the spacelike infinity is located at $x=y=-1$. 
The coordinates $(x,\phi)$ parametrize the two-sphere $S^2$ and $\psi$ 
parametrizes the circle $S^1$. 
One recovers the Emparan-Reall black ring by setting $\nu=0$, and 
the line $\nu=\lambda^2/4$ represents
the sequence of extremal black rings 
(see Fig.~\ref{Fig:parameter}).

%===========<Figure1>============%
%
%
%=================================%

The mass and angular momenta are
\begin{equation}
\label{physical parameters start}
M=\frac{3\pi R^2\lambda}{1-\lambda+\nu},
\qquad
J_\phi=\frac{4\pi R^3\lambda\sqrt{\nu}\sqrt{(1+\nu)^2-\lambda^2}}{(1-\nu)^2(1-\lambda+\nu)},
\end{equation}
\begin{equation}
J_\psi=\frac{2\pi R^3\lambda(1+\lambda-6\nu+\lambda\nu+\nu^2)\sqrt{(1+\nu)^2-\lambda^2}}{(1-\nu)^2(1-\lambda+\nu)^2}.
\end{equation}
The angular velocities, the area, and the surface gravity 
of the horizon are written as~\cite{Elvang:2007hs}
\begin{equation}
\Omega_\phi=\frac{\lambda(1+\nu)-(1-\nu)\sqrt{\lambda^2-4\nu}}{4R\lambda\sqrt{\nu}}\sqrt{\frac{1+\nu-\lambda}{1+\nu+\lambda}}, 
\qquad
\Omega_\psi=\frac{1}{2R}\sqrt{\frac{1+\nu-\lambda}{1+\nu+\lambda}},
\end{equation}
\begin{equation}
\label{physical parameters end}
A_H=\frac{32\pi^2R^3\lambda(1+\nu+\lambda)}{(1-\nu)^2(y_h^{-1}-y_h)},
\qquad
\kappa=\frac{(y_h^{-1}-y_h)(1-\nu)\sqrt{\lambda^2-4\nu}}{4R\lambda(1+\nu+\lambda)}.
\end{equation}

\subsection{Thin ring limit}
We consider a thin ring limit $\lambda\to 0$ 
where the ratio of the $S^2$ radius 
to the $S^1$ radius becomes very small. 
Here, we have to take care of the fact that this 
limit depends on the path to the point $\lambda=\nu=0$. 
For example, taking the limit $\lambda\to 0$ on the line $\nu=0$
gives a boosted Schwarzschild string, 
while taking the limit $\lambda\to 0$ on the extremal line $\nu = \lambda^2/4$
should result in an extremal Kerr black string. 
Therefore, $\lambda = \nu =0$ is a degenerate point, and
in order to resolve this degeneracy,
we introduce a new parameter $a_\ast$ as
\begin{equation}
\label{parameters}
\nu=\frac14 a_\ast^2\lambda^2,
\end{equation}
and consider a limit $\lambda\to 0$ on the line of a fixed $a_\ast$
(see Fig.~\ref{Fig:parameter}). Also, in order to obtain a well-defined
limit, we introduce 
\begin{equation}
M_K = \frac{1}{\sqrt{2}}\lambda R,
\end{equation}
and fix $M_K$ in taking this limit.

We introduce the new coordinates $r$, $z$ and $\theta$ as
\begin{equation}
\label{coordinates}
y=-\frac{\sqrt{2}R}{r}, 
\qquad 
\psi=-\frac{z}{\sqrt{2}R}, 
\qquad 
x=\cos\theta,
\end{equation}
and collect the leading-order term of each metric component
with respect to $\lambda$.
Then, 
the black ring solution is reduced to the so-called 
boosted Kerr string solution
\begin{multline}
\label{BKBS}
ds^2=-\left(1-\frac{2M_Kr\cosh^2\sigma}{\rho^2}\right)dt^2+\frac{2M_Kr\sinh2\sigma}{\rho^2}dtdz
     +\left(1+\frac{2M_Kr\sinh^2\sigma}{\rho^2}\right)dz^2
\\   +\frac{\rho^2}{\Delta}dr^2+\rho^2d\theta^2
     +\frac{\left(r^2+a^2\right)^2-\Delta a^2\sin^2\theta}{\rho^2}\sin^2\theta d\phi^2
\\   -\frac{4M_Kr\cosh\sigma}{\rho^2}a\sin^2\theta dtd\phi-\frac{4M_Kr\sinh\sigma}{\rho^2}a\sin^2\theta dzd\phi,
\end{multline}
where $\rho^2=r^2+a^2\cos^2\theta$ and $\Delta=r^2-2M_Kr+a^2$, and
$a$ is defined by $a:= M_Ka_\ast$. 
Since $M_K$ and $a$ correspond to the mass and rotational parameter 
of a four-dimensional Kerr black hole, respectively, 
$a_\ast$ represents the nondimensional 
rotation parameter along $S^2$ direction. 
$\sigma:=\mathrm{arctanh}(1/\sqrt{2})$ is the boost parameter. 
Although the boost parameter can take any value for a general boosted Kerr string, 
it is restricted to this value for the thin limit of a black ring.
The event horizon is located at $r=r_+$ where
$r_\pm:=M_K\pm\sqrt{M_K^2-a^2}=M_K (1 \pm \sqrt{1-a_*^2})$.

In studying the evaporation of a black ring, 
we use this boosted Kerr string solution in the following sense.
We consider the situation where $\lambda$ is very small,
and do not take the exact limit.
Then, in the neighborhood of the black ring, the spacetime
metric can be well approximated by the boosted Kerr string solution.
For this reason, the value of $R$ is not infinite 
in our analysis although it is very large compared to $M_K$. 
The relative error in this approximation is $O(\lambda)$ compared to the leading
order in the following analyses.

In this thin-limit approximation, the physical quantities in 
Eqs.~\eqref{physical parameters start}--\eqref{physical parameters end} 
are expressed in terms of $R, M_K$ and $a$ (or $a_*$) as
\begin{equation}
\label{relation1}
M \simeq 3\sqrt{2} \pi R M_K, \qquad
J_\psi \simeq 2\sqrt{2} \pi R^2 M_K, \qquad
J_\phi \simeq 4 \pi a_\ast R M_K^2.
\end{equation}
\begin{equation}
\label{Omega_leading_order}
\Omega_\phi \simeq \frac{a}{2M_Kr_+}\frac{1}{\cosh\sigma},
\qquad
\Omega_\psi \simeq \frac{1}{2R},
\end{equation}
\begin{equation}
\label{Eq:kappa}
A_H \simeq 16 \pi^2 r_+ M_K R,
\qquad
\kappa \simeq \frac{r_+-r_-}{4 M_K r_+}\frac{1}{\cosh\sigma}.
\end{equation}
The inverse relations of Eq.~\eqref{relation1} are 
\begin{equation}
\label{relation2}
R \simeq \frac{3J_\psi}{2M}, \qquad
M_K \simeq \frac{\sqrt{2}}{9\pi}\frac{M^2}{J_\psi}, \qquad
a_\ast \simeq \frac{27\pi}{4}\frac{J_\psi J_\phi}{M^3}, 
\end{equation}
and $\lambda$ is expressed as
\begin{equation}
\lambda = \frac{\sqrt{2}M_K}{R}  \simeq \frac{4}{27\pi}\frac{M^3}{J_\psi^2}.
\label{Eq:lambda}
\end{equation}
Because $\lambda^{-1/2}$ is proportional to 
the angular momentum $J_\psi$ normalized by mass $M$, 
it can be interpreted
as the nondimensional rotation parameter along $S^1$. 
At the same time, Eq.~\eqref{Eq:lambda} also means that 
$\lambda$ gives the order of the ratio of the $S^2$ radius 
to the $S^1$ radius. Therefore, $\lambda$ is interpreted as an
indicator for the ``thickness'' of the black ring.
In this paper, we call $\lambda$ the thickness parameter.

\subsection{Effect of boost}
Note that $\Omega_\phi$ in Eq.~\eqref{Omega_leading_order} is equal to   
the angular velocity defined by the Killing generator $\xi$
of the horizon of the boosted Kerr string\footnote{
Our expression of $\Omega_\phi$ does not agree with that of Ref.~\cite{Dias:2006zv} 
because the definition is different.},
\begin{equation}
\xi = \partial_t + \Omega_\phi\partial_\phi + V \partial_z 
\end{equation}
with
\begin{equation}
V=\tanh\sigma,
\label{Eq:linear_velocity}
\end{equation}
and $\kappa$ in Eq.~\eqref{Eq:kappa} is identical to the surface gravity of the horizon calculated with $\xi$. 
For a later convenience, it is useful to compare $\Omega_\phi$ and $\kappa$ 
with the angular velocity and the surface gravity of the horizon of the unboosted Kerr string. 
In the following, quantities in the unboosted system are indicated
by prime $(\ {}^\prime\ )$. In the unboosted system, 
the Killing generator of the horizon is
$\xi^\prime = \partial_{t^\prime} + \Omega_{\phi}^\prime \partial_\phi$, 
and $\Omega_\phi^\prime$ and $\kappa^\prime$ 
calculated from $\xi^\prime$ are
\begin{equation}
\Omega_\phi^\prime = \frac{a}{2M_Kr_+},
\qquad
\kappa^\prime = \frac{r_+-r_-}{4 M_Kr_+}.
\label{Omega-kappa-prime}
\end{equation}
There is a deference in the quantities 
of the boosted and unboosted systems by a factor of $1/\cosh\sigma$.
This is understood as the effect of time delay in the
Lorentz boost.

%
%======================================%
%<<<<<<<<<<<< SECTION III  >>>>>>>>>>>>>>%
%======================================%
%
\section{Formulation}
In this section, we formulate the time evolution of mass and angular momenta 
of a thin black ring via Hawking radiation, 
by approximating the evolution of a scalar field in the black ring spacetime 
by that in a boosted Kerr string spacetime. 

\subsection{Emission rate}
\label{section:Emission rate}
The evolution of a scalar field is governed by the Klein-Gordon equation in curved spacetime 
\begin{equation}
\label{KG}
\left(-g\right)^{-1/2}\partial_\mu\left(\sqrt{-g}g^{\mu\nu}\partial_\nu\Phi\right)=0,
\end{equation}
where $g$ is the determinant of its metric.

To quantize the field, we need to expand it in terms of the eigenmodes for $\Phi$, which can be written in the black ring background as
\begin{equation}
\Phi= e^{-i\omega t}e^{im\phi} e^{in\psi}\Psi(x,y),
\end{equation}
where $\omega$, $m$ and $n$ are the eigenvalues for the Killing vector fields $\partial_t$, $\partial_\psi$ and $\partial_\phi$, respectively.  
By inserting this expression into Eq.~\eqref{KG}, 
we obtain a second-order elliptic equation for $\Psi(x,y)$ 
in the $(x,y)$ plane. 
This equation has a discrete series of regular solutions labeled by an 
integer $\ell$. 
In the Schwarzschild string limit with $J_\phi=0$, 
this series of solutions become proportional to the associate 
Legendre functions $P^m_\ell(x)$. 
Thus, the mode functions are labeled by the four parameters 
$(\omega,\ell,m,n)$ in which $\ell$, $m$ and $n$ take integer values.

In the case of a nonrotating black hole, 
the expected number of particles emitted per unit time for each mode is proportional to black body radiation: 
\begin{equation}
\langle N_s \rangle \propto \frac{1}{e^{\omega/T_s}-1},
\end{equation}
where $\omega$ is the energy of a scalar particle in the background of the nonrotating black hole and $T_s$ is the temperature of the horizon. 
Here, the temperature is expressed as $T_s = \kappa_s /2\pi$ 
in terms of the surface gravity $\kappa_s$ of the horizon. 

In the rotating case, we have to replace $\omega$ by the energy of the mode with respect to the null geodesic generator of the black hole horizons 
because the mode function behaves as $\exp( -i\omega_* u_\pm)$ in the coordinates that are regular around the black hole horizon, 
where $u_\pm$ is the advanced time/retarded time around the horizon. 
In general, this null geodesic generator $\xi$ can be written as 
$\xi=\partial_t +\sum_j \Omega_j \partial_{\phi^j}$ in terms of the time translation Killing vector $\partial_t$ 
and the rotational Killing vectors $\partial_{\phi^j}$.  
From this, it follows that $\omega_*$ for the mode $\propto \exp(-i\omega t + i\sum_j m_j\phi^j)$ is expressed as
\begin{equation}
\omega_*= \omega- \sum_j  m^j \Omega_j.
\end{equation}
Hence, the expected number of particles emitted per unit time for each mode from the black ring is given by
\begin{equation}
\langle N_{\rm BR}\rangle 
= \frac{\Gamma^{\rm (BR)}_{{\ell} m n}(\omega)}
{e^{(\omega-n\Omega_\psi-m\Omega_\phi)/T_{\rm BR}}-1}, 
\end{equation}
where $T_{\rm BR}$ is the temperature of the horizon and 
$\Gamma^{\rm (BR)}_{{\ell} m n}(\omega)$ is the greybody factor,
which is identical to the absorption probability of the incoming wave of the corresponding modes.
This determines the emission rates of the total mass $M$ and angular momenta $J_\psi$ and $J_\phi$ as
\begin{equation}
\label{emission}
-\frac{d}{dt}
\left(
\begin{array}{ccc}
M \\
J_\psi \\
J_\phi \\
\end{array} 
\right)
=\frac{1}{2\pi}\sum_{\ell,m,n}\int^\infty_0 d\omega\langle N_{\rm BR}\rangle
\left(
\begin{array}{ccc}
\omega \\
n \\
m \\
\end{array} 
\right),
\end{equation}
where the summation is taken over all modes.
Note that in this expression, it is difficult to estimate the greybody factor generally 
because we cannot separate the coordinates $x$ and $y$, 
and two-dimensional numerical calculations are required 
to determine the energy eigenvalues and corresponding eigenmodes.

In order to circumvent this difficulty, 
we consider the situation where the mode functions
can be approximately evaluated: 
a black string limit discussed above. 
For the boosted black string \eqref{BKBS}, 
we can separate the wave equation, and therefore, 
we approximate the evolution of a scalar field in a black ring spacetime 
by that in a boosted Kerr string spacetime.
In this situation, the variables can be separated as 
\begin{equation}
\label{separation}
\Phi=e^{-i\omega t}R(r)e^{-ikz}e^{im\phi}S^m_\ell(\theta), 
\end{equation}
where $S^m_\ell(\theta)$ is the spheroidal harmonic function.
From the coordinate transformation \eqref{coordinates}, 
$n$ of a black ring and $k$ of a boosted black string are related by
\begin{equation}
\label{relation_n_and_k}
n=2kR\tanh\sigma. 
\end{equation}
The expected number of particles emitted per unit time per mode is given by
\begin{equation}
\langle N_{\rm BBS}\rangle
=\frac{\Gamma_{\ell m n}(\omega)}{e^{(\omega-kV-m\Omega_\phi)/T}-1},
\end{equation}
where $T=\kappa/2\pi$ is the temperature of the horizon 
with $\kappa$ in Eq.~\eqref{Eq:kappa}, and 
$V$ is the linear velocity introduced in Eq.~\eqref{Eq:linear_velocity}.
$\Gamma_{\ell m n}(\omega)$ is the greybody factor of 
the boosted Kerr string spacetime.
We evaluate the emission rates \eqref{emission} 
by using $\langle N_{\rm BBS} \rangle$ instead of $\langle N_{\rm BR} \rangle$.

\subsection{Simplification}
\label{Sec:Simplification}
We normalize all quantities by the mass density $M_K$, 
\begin{equation}
\label{Eq:def_tildeomega_etc}
\tilde{\omega} = M_K\omega, \qquad
\tilde{k} = M_K k, \qquad
\tilde{\Omega}_\phi = M_K \Omega_\phi, \qquad
\tilde{T} = M_K T= \frac{M_K\kappa}{2\pi}.
\end{equation}
We change the order of summations over $\ell$ and $m$ as
\begin{equation}
\sum_{\ell=0}^{\infty} \sum_{m=-\ell}^{\ell} = \sum_{m=-\infty}^{\infty} \sum_{\ell=|m|}^{\infty},
\end{equation}
and introduce
\begin{equation}
g^{(m)}(\tilde{\omega},\tilde{k}) := \sum_{\ell=|m|}^{\infty}\Gamma_{\ell m n}(\tilde{\omega}),
\label{Eq:def_gm}
\end{equation}
where $\tilde{k}$ and $n$ are related to each other by Eq.~\eqref{relation_n_and_k}. 
Then, the emission rates can be written as
\begin{equation}
-\frac{d}{dt}
\left(
\begin{array}{ccc}
M \\
J_\psi \\
J_\phi \\
\end{array} 
\right)
=\frac{1}{2\pi M_K} \sum_{m=-\infty}^{\infty} \sum_{n=-\infty}^{\infty} \int^{\infty}_{|\tilde{k}|}d\tilde{\omega}
 \frac{g^{(m)}(\tilde{\omega},\tilde{k})}{e^{(\tilde{\omega}-\tilde{k}V-m\tilde{\Omega}_\phi)/\tilde{T}}-1}
\left(
\begin{array}{ccc}
\tilde{\omega} / M_K \\
n \\
m \\
\end{array} 
\right).
\label{Eq:simplification1}
\end{equation}
Here, the lower limit of the integral is $|\tilde{k}|$ 
because the modes with their energy $\omega<|\tilde{k}|$ are gravitationally
bounded and do not escape to infinity. 
Because the spectral density of  $\tilde k$ is very large, $O(1/\lambda)$, 
the summation over $n$ can be replaced by the integral: 
\begin{equation}
\sum_n \to \int dn = \frac{2R\tanh\sigma}{M_K}\int d\tilde{k}.
\end{equation}
The relative error produced by this replacement 
is $O(\lambda)$ and negligible in our thin-limit approximation.
We obtain 
\begin{equation}
-\frac{dM}{dt} = \frac{R}{\sqrt{2}\pi M_K^3}
\sum_{m=-\infty}^\infty
\int_{-\infty}^\infty d\tilde{k}
\int_{|\tilde{k}|}^{\infty}d\tilde{\omega}
\frac{\tilde{\omega}
g^{(m)}(\tilde{\omega}, \tilde{k})}
{e^{(\tilde{\omega}-\tilde{k}V-m\tilde{\Omega}_\phi)/\tilde{T}}-1},
\label{Energy-RadiationRate-without-prime}
\end{equation}
\begin{equation}
-\frac{dJ_\psi}{dt} 
= \frac{R^2}{\pi M_K^3}
\sum_{m=-\infty}^\infty
\int_{-\infty}^\infty d\tilde{k}
\int_{|\tilde{k}|}^{\infty}d\tilde{\omega}
\frac{\tilde{k}
g^{(m)}(\tilde{\omega}, \tilde{k})}
{e^{(\tilde{\omega}-\tilde{k}V-m\tilde{\Omega}_\phi)/\tilde{T}}-1}.
\label{AMpsi-RadiationRate-without-prime}
\end{equation}
\begin{equation}
-\frac{dJ_\phi}{dt} 
= \frac{R}{\sqrt{2}\pi M_K^2}
\sum_{m=-\infty}^\infty
\int_{-\infty}^\infty d\tilde{k}
\int_{|\tilde{k}|}^{\infty}d\tilde{\omega}
\frac{m
g^{(m)}(\tilde{\omega}, \tilde{k})}
{e^{(\tilde{\omega}-\tilde{k}V-m\tilde{\Omega}_\phi)/\tilde{T}}-1}.
\label{AMphi-RadiationRate-without-prime}
\end{equation}  
The integral in each formula can be further simplified 
if we perform the transformation from $(\tilde{\omega},\tilde{k})$ to 
$(\tilde{\omega}^\prime, \tilde{k}^\prime)$, 
\begin{equation}
\label{Lorentz tr}
\tilde{\omega} = \tilde{\omega}^\prime \cosh\sigma + \tilde{k}^\prime \sinh\sigma,
\qquad
\tilde{k} = \tilde{\omega}^\prime \sinh\sigma + \tilde{k}^\prime \cosh\sigma. 
\end{equation}
Substituting these formulas with $\cosh\sigma=\sqrt{2}$ and $\sinh\sigma=1$ and 
rewriting $M_K$ and $R$ by $M$, $J_\psi$ and $J_\phi$ using Eq.~\eqref{relation2},
we obtain the simplified expression of the evolution:
\begin{equation}
\label{time evolution of M}
-\frac{1}{M}\frac{dM}{dt} = 
2F(a_\ast)\frac{J_\psi^4}{M^8},
\end{equation}
\begin{equation}
\label{time evolution of Jpsi}
-\frac{1}{J_\psi}\frac{dJ_\psi}{dt} = 
3F(a_\ast)\frac{J_\psi^4}{M^8},
\end{equation}
\begin{equation}
\label{time evolution of Jphi}
-\frac{1}{J_\phi}\frac{dJ_\phi}{dt} = 
G(a_\ast)\frac{J_\psi^3}{J_\phi M^5},
\end{equation}
with
\begin{equation}
F(a_\ast) := 
\frac{3^7\pi^2}{2^3\sqrt{2}} 
\sum_{m=-\infty}^{\infty} I_1^{(m)}(a_\ast),
\label{Eq:def_Fa}
\end{equation}
\begin{equation}
\label{Eq:def_Ga}
G(a_\ast) := 
\frac{3^5\pi}{2^2\sqrt{2}}
\sum_{m=-\infty}^{\infty} I_2^{(m)}(a_\ast).
\end{equation}
Here, we defined
\begin{equation}
I_1^{(m)}(a_\ast) := \int^{\infty}_{-\infty}d\tilde{k}^\prime \int^{\infty}_{|\tilde{k}^\prime|}d\tilde{\omega}^\prime
             \frac{\tilde{\omega}^\prime g^{\prime (m)}(\tilde{\omega}^\prime,\tilde{k}^\prime)}{e^{(\tilde{\omega}^\prime - m\tilde{\Omega}^\prime_\phi)/\tilde{T}^\prime} - 1},
\label{Eq:def_I1m}
\end{equation}
\begin{equation}
I_2^{(m)}(a_\ast) := \int^{\infty}_{-\infty}d\tilde{k}^\prime \int^{\infty}_{|\tilde{k}^\prime|}d\tilde{\omega}^\prime
             \frac{m g^{\prime (m)}(\tilde{\omega}^\prime,\tilde{k}^\prime)}{e^{(\tilde{\omega}^\prime - m\tilde{\Omega}^\prime_\phi)/\tilde{T}^\prime} - 1},
\end{equation}
with $g^{\prime (m)}(\tilde{\omega}^\prime,\tilde{k}^\prime) = g^{(m)}(\tilde{\omega},\tilde{k})$. 
In the same manner as Eq.~\eqref{Eq:def_tildeomega_etc}, 
we normalized the quantities 
${\Omega}_{\phi}^\prime$ and ${T}^\prime=\kappa^\prime/2\pi$ 
of the unboosted system in Eq.~\eqref{Omega-kappa-prime} as 
$\tilde{\Omega}_{\phi}^\prime:= M_K \Omega_\phi^\prime$ and 
$\tilde{T}^\prime := M_K T^\prime$. Their explicit formulas are
\begin{equation}
\tilde{\Omega}_\phi^\prime = \frac{(1/2)a_*}{1+\sqrt{1-a_*^2}}, \qquad
\tilde{T}^\prime =  \frac{(1/4\pi)\sqrt{1-a_*^2}}{1+\sqrt{1-a_*^2}},
\end{equation}
and depend only on $a_*$. 
Note that we used the fact that
the integrals of terms proportional to $\tilde{k}^\prime$ vanish
because the greybody factor is an even function 
of $\tilde{k}^\prime$ (see Eqs.~\eqref{Eq:radial_equation} and 
\eqref{Eq:radial_potential} of the next subsection), and thus, 
such terms are odd functions of $\tilde{k}^\prime$.
From Eqs.~\eqref{time evolution of M}--\eqref{time evolution of Jphi} 
and Eq.~\eqref{relation2}, the equation for $a_\ast$ is derived as
\begin{equation}
\label{time evolution of a}
-\frac{1}{a_\ast}\frac{da_\ast}{dt} = 
3H(a_\ast)\frac{J_\psi^4}{M^8},
\end{equation}
where
\begin{equation}
\label{Eq:def_Ha}
H(a_\ast) := \frac{9\pi}{4}
\frac{G(a_\ast)}{a_\ast} - F(a_\ast).
\end{equation}
Therefore, the time evolution of a thin black ring by the Hawking radiation is 
determined by the equations for $M$, $J_\psi$ and $a_\ast$, 
that is, Eqs.~\eqref{time evolution of M}, \eqref{time evolution of Jpsi} 
and \eqref{time evolution of a}. 
The remaining task is to calculate the greybody factors and obtain $F(a_\ast)$ and $H(a_\ast)$ of Eqs.~\eqref{Eq:def_Fa} and \eqref{Eq:def_Ha} numerically.

\subsection{Greybody factor}
In the following, we discuss the greybody factor 
for a massless scalar field in a boosted Kerr string spacetime. 
Substituting the ansatz \eqref{separation} into 
the Klein-Gordon equation \eqref{KG} 
in the background of the boosted Kerr string \eqref{BKBS}, 
we get the following angular and radial wave equations 
for $S^m_\ell(\theta)$ and $R(r)$ \cite{Dias:2006zv}:
\begin{equation}
\label{spheroidal}
0=\frac{1}{\sin\theta}\partial_\theta\left(\sin\theta\partial_\theta S^m_\ell\right)
    +\left[a^2\left(\omega^2-k^2\right)\cos^2\theta-\frac{m^2}{\sin^2\theta}+\lambda_{\ell m}\right]S^m_\ell,
\end{equation}
\begin{multline}
\label{radial}
0=\Delta\partial_r\left(\Delta\partial_rR\right)
   -\Delta\left[k^2r^2+a^2\omega^2-2\omega ma\cosh\sigma+\lambda_{\ell m}\right]R
\\+\Big[
\left[\omega\left(r^2+a^2\right)-ma\cosh\sigma\right]^2
       +2M_Kr\left(r^2+a^2\right)\cosh^2\sigma\left(\omega-k\tanh\sigma\right)^2
\\       -2M_Kr\left(r^2+a^2\right)\omega^2
       -m^2a^2\sinh^2\sigma+4kmaM_Kr\sinh\sigma\Big]R,
\end{multline}
where $\lambda_{\ell m}$ is the separation constant 
which is determined as an eigenvalue of \eqref{spheroidal}. 
For small $a^2\left(\omega^2-k^2\right)$, the eigenvalues 
associated with the spheroidal wave functions $S^m_\ell$ are 
$\lambda_{\ell m}=\ell\left(\ell+1\right)+{\cal O}
\left(a^2\left(\omega^2-k^2\right)\right)$~\cite{Teukolsky:1973ha}.

As we mentioned in Sec.~\ref{section:Emission rate}, the greybody factor is calculated 
as the absorption probability of the incoming waves of the corresponding mode.
With the tortoise coordinate $r_\ast$ and the new wave function $u$ defined by
\begin{equation}
dr_\ast=\frac{r^2+a^2}{\Delta}dr, \qquad R=\frac{1}{\sqrt{r^2+a^2}}u,
\end{equation}
the radial wave equation \eqref{radial} can be rewritten as the following equation of Schr\"odinger type:
\begin{equation}
\left[\frac{d^2}{dr_\ast^2}+{\omega^\prime}^2-V(r) \right]u=0,
\label{Eq:radial_equation}
\end{equation}
where $V(r)$ is the effective potential
\begin{equation}
V(r)=\frac{\Delta\left\{2M_Kr(r^2-2a^2)+a^2(r^2+a^2)\right\}}{(r^2+a^2)^4}
       +\frac{\Delta({\omega^\prime}^2a^2+\lambda_{\ell m}+{k^\prime}^2r^2)+4m\omega^\prime M_Kar-m^2a^2}{(r^2+a^2)^2}.
\label{Eq:radial_potential}
\end{equation}
Here, the quantities in the unboosted frame, $\omega^\prime$ and $k^\prime$,
were introduced in the same manner as Eq.~\eqref{Lorentz tr}.
Note that Eqs.~\eqref{Eq:radial_equation} and \eqref{Eq:radial_potential}
have the same form as the equations for a massive scalar field with 
angular frequency $\omega^\prime $ and mass $|k^\prime|$
in a four-dimensional Kerr spacetime. 
As the boundary condition, we impose the ingoing condition at the horizon. 
Then, $u$ behaves as
\begin{equation}
u(r_\ast)\sim
\begin{cases}
  &e^{-i\omega_+^\prime r_\ast} \hspace{40mm} \text{at} \qquad r \to r_+,
\\&A_{in}e^{-i\omega_{\infty}^\prime r_\ast}+A_{out}e^{i\omega_{\infty}^\prime r_\ast} \hspace{10mm} \text{at} \qquad r \to \infty.
\end{cases}
\end{equation}
Here $\omega_+^\prime:=\omega^\prime-m\Omega_\phi^\prime$ 
and $\omega_\infty^\prime:=\sqrt{{\omega^\prime}^2-{k^\prime}^2}$.
The greybody factor is written as
\begin{equation}
\Gamma^\prime_{\ell m}(\omega^\prime, k^\prime)
=
1-\left|\frac{A_{out}}{A_{in}}\right|^2,
\end{equation}
which has to be evaluated numerically. 

\begin{figure}[tb]
\begin{tabular}{cc}
\includegraphics[width=0.5\textwidth]{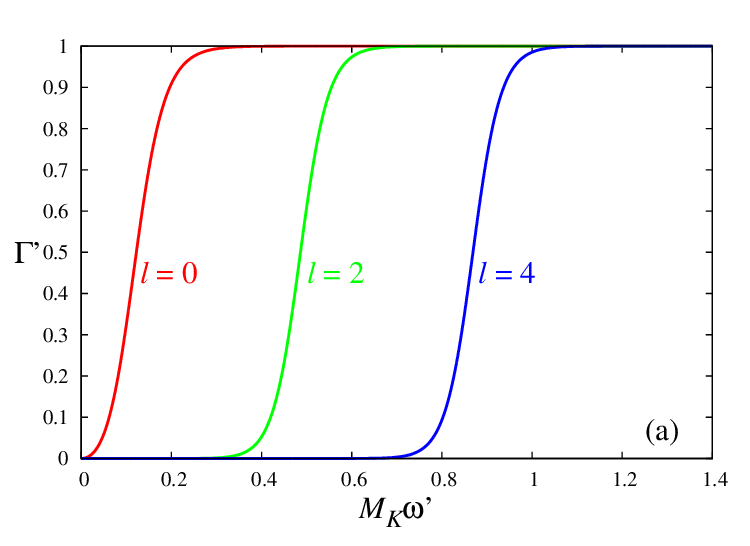}
\includegraphics[width=0.5\textwidth]{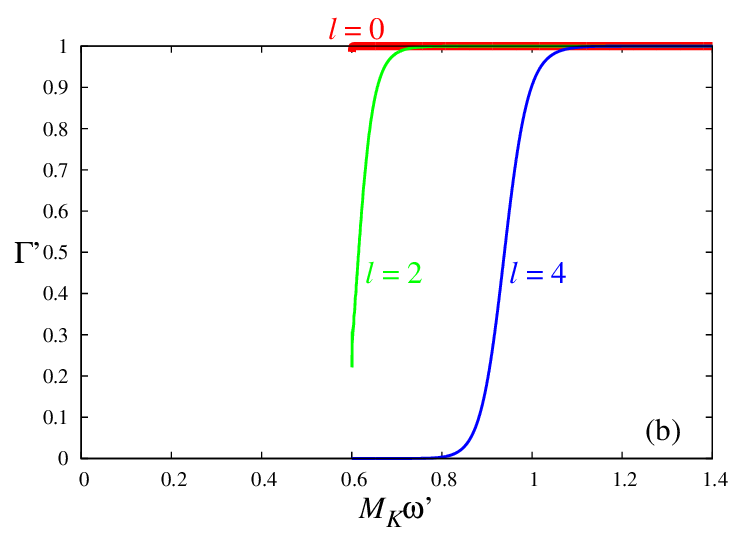}
\end{tabular}
\caption{
Greybody factors $\Gamma^\prime$ as functions of $M_K\omega^\prime$ 
for the modes $\ell=0$, $2$, and $4$ in the case $a_*=0$ 
for (a) $M_Kk^\prime=0$ (left panel) and (b) $M_Kk^\prime=0.6$ (right panel).
In the right panel, the data for $M_K\omega^\prime \ge 0.6005$ are
plotted.
}
\label{Fig:gf}
\end{figure}

In this paper, we perform numerical calculations of the greybody factor 
in the case of the Emparan-Reall black ring, i.e., $a=0$.
In this case, $S^m_\ell(\theta) e^{im\phi}$ becomes 
the spherical harmonic function $Y_{\ell m}(\theta,\phi)$ 
and the greybody factor is independent of $m$, and therefore,
the calculation becomes much simpler compared to the case $a\neq 0$. 
We developed a code to calculate the greybody factor in this situation. 
The left and right panels of Fig.~\ref{Fig:gf} 
show the behaviors of the greybody factors 
of the first three even $\ell$ numbers 
for $\tilde{k}^\prime=0$ and $0.6$, respectively.
Our result is in good agreement with the 
analytic approximate formula for low-frequency waves 
in Ref.~\cite{Unruh:1976fm} (see also \cite{Jung:2004nh,Kanti:2010mk}). 

%===========<Figure2>============%
%
%
%=================================%

Note that if we take the limit $\tilde{\omega}^\prime\to \tilde{k}^\prime$
for $\tilde{k}^\prime\neq 0$,
the greybody factor is expected to approach a nonzero finite 
value.\footnote{
This behavior has been claimed in Ref.~\cite{Unruh:1976fm} 
and we have also checked it using an analytic toy model.
} 
Obtaining these values by numerical calculation
is rather difficult because the greybody factors have to be 
evaluated at a very distant position $r/M\gg 1/v^{\prime 2}$ with 
$v^\prime=\sqrt{1-(k^\prime/\omega^\prime)^2}$. 
Although these values are left uncertain in our calculation, 
we have checked that the error caused by this uncertainty is small. 
For example, the error in $F(0)$ of Eq.~\eqref{Eq:result_F0} 
is smaller than 0.1\%.

%
%======================================%
%<<<<<<<<<<<< SECTION IV  >>>>>>>>>>>>>>%
%======================================%
%
\section{Evolution by Evaporation}

In this section, we discuss general features of 
the time evolution of the evaporating Pomeransky-Sen'kov black ring. 
Then, we focus attention to the case of the Emparan-Reall black ring, 
and derive semi-analytic solutions of the time evolution
using the numerical results of the greybody factors.

\subsection{Evolution of Pomeransky-Sen'kov black rings}

First, we discuss general features that does not depend on the
details of the greybody factor. 
From Eqs.~\eqref{time evolution of M} and \eqref{time evolution of Jpsi}, 
the following relation can be immediately found:
\begin{equation}
\label{result S1}
\frac{J_\psi^2}{M^3}= \text{const.}
\end{equation}
From Eq.~\eqref{Eq:lambda}, this means that the thickness parameter 
$\lambda$ does not change,
\begin{equation}
\label{lambda}
\lambda(t) \equiv \lambda(0).
\end{equation}
Therefore, a Pomeransky-Sen'kov black ring 
evaporates without changing the initial value of the 
nondimensional rotation parameter along $S^1$.

Next, let us assume $a_*>0$ 
and consider how to solve the evolution equations for $a_*(t)$ and $M(t)$.
Eliminating $M$ and $J_\psi$ from Eqs.~\eqref{time evolution of M} and \eqref{time evolution of a} using Eq.~\eqref{result S1}, 
we obtain
\begin{equation}
\frac{d}{dt}\left(\frac{3a_*H(a_*)}{da_*/dt}\right)=4F(a_*).
\end{equation}
In principle, this equation can be solved at least numerically to yield $a_\ast(t)$ 
once $H(a_*)$ and $F(a_*)$ are specified. 
Then, from Eqs.~\eqref{time evolution of M} and \eqref{time evolution of a}, 
the time evolution of $M(t)$ is formally given by
\begin{equation}
\label{M(t) of PS}
M(t)=M(0)\exp\left[ 
\frac{2}{3}
\int_{a_\ast(0)}^{a_\ast(t)}\frac{F(a_\ast)}{a_\ast H(a_\ast)}da_\ast \right].
\end{equation}

We point out that the behavior of $H(a_*)$ is crucial for the evolution
of $a_*$. This function is analogous to $h(a_*)$
of Ref.~\cite{Chambers:1997ai} where evolution of a four-dimensional
Kerr black hole was investigated: The value of $a_*$ increases (decreases) 
if $H(a_*)$ is negative (positive). If $H(a_*)$ crosses 
zero from negative to positive at some value $a_* = a_*^{(c)}$ 
similarly to Fig.~3 of Ref.~\cite{Chambers:1997ai}, 
the black ring inevitably evolves to the state with $a_*^{(c)}$. 
Therefore, numerical calculation of $H(a_*)$ is very interesting, but is left for future work. 
In the present paper, we only study what happens in the case of an Emparan-Reall black ring.

\subsection{Evolution of Emparan-Reall black rings}
\label{Sec:Evolution_of_ER_BR}

In the case of the Emparan-Reall black ring, 
$J_\phi(t) \equiv 0$ and $a_\ast \equiv 0$, 
and Eqs.~\eqref{time evolution of M} and 
\eqref{time evolution of Jpsi} can be solved analytically:
\begin{equation}
\label{M_of_ER}
M(t)=M(0)\left( 1 - 4F(0)\frac{J_\psi(0)^4}{M(0)^8}t \right)^{1/2},
\end{equation}
\begin{equation}
\label{Jpsi_of_ER}
J_\psi(t)=J_\psi(0)\left( 1 - 4F(0)\frac{J_\psi(0)^4}{M(0)^8}t \right)^{3/4}.
\end{equation}
Therefore, our remaining task is to determine $F(0)$ numerically. 

As discussed in Sec.~\ref{Sec:Simplification}, 
basically we have to calculate
Eqs.~\eqref{Eq:def_gm}, \eqref{Eq:def_I1m}, and \eqref{Eq:def_Fa}, and
in those equations, the summation over $m$ was
taken at the last step.
But in the case of $a_*=0$ considered
here, it is better to take the summation with respect to $m$
in advance because the integrand does not depend on $m$.
For this reason,  we calculate summation of the greybody factors as
\begin{equation}
g^\prime(\tilde{\omega}^\prime, \tilde{k}^\prime)
:=\sum_{\ell=0}^{\infty}\sum_{m=-\ell}^{\ell}
\Gamma^\prime_{\ell m}(\tilde{\omega}^\prime,\tilde{k}^\prime)
=\sum_{\ell=0}^{\infty}
(2\ell +1)\Gamma^\prime_{\ell}(\tilde{\omega}^\prime,\tilde{k}^\prime).
\label{Eq:def_g}
\end{equation}
Then, $F(0)$ is given by
\begin{equation}
F(0) = \frac{3^7\pi^2}{2^3\sqrt{2}}I_1,
\qquad
I_1:=\int_{-\infty}^{\infty}d\tilde{k}^\prime
\int_{|\tilde{k}^\prime|}^{\infty}d\tilde{\omega}^\prime
\frac{\tilde{\omega}^{\prime}
g^\prime(\tilde{\omega}^\prime, \tilde{k}^\prime)}
{e^{\tilde{\omega}^\prime/\tilde{T}^\prime}-1}.
\label{Eq:F0_ER}
\end{equation}

The integrations of $I_1$ were executed with the Simpson's method. 
The domain of integration of $I_1$ was made finite 
by discarding the region where the integrand is sufficiently small. 
We therefore set the upper limit of integration with respect to $\tilde{\omega}^\prime$ to be $0.75$.
We took summation with respect to $\ell$ up to $\ell=5$, 
because the contributions from the modes $\ell > 5$ 
turned out to be negligible. 
This is because the potential walls for $\ell > 5$ are 
so high that they reflect almost of all waves. 
In this manner, $F(0)$ is determined as 
\begin{equation}
F(0)\simeq 0.239. 
\label{Eq:result_F0}
\end{equation}
As a check, we compute $F(0)$ using the DeWitt approximation \cite{DeWitt:1975} 
in Appendix \ref{Appendix_A}. 
The two results agree well, and therefore, our numerical result is reliable.

As shown in Eq.~\eqref{lambda}, 
the nondimensional rotation parameter $J_\psi^2/M^3$ 
is held fixed throughout the evolution, and this also indicates
the constancy of the thickness parameter $\lambda$. 
Because $a_*\equiv 0$, 
the Emparan-Reall black ring evaporates keeping 
similarity to its initial shape: 
The black ring at any time 
can be obtained by uniformly scaling its initial configuration. 
The scaling factor $C(t)$ can be found by deriving the
time evolution of the ring radius $R$ as
\begin{equation}
C(t) := \frac{R(t)}{R(0)} = \left( 1 - 4F(0)\frac{J_\psi(0)^4}{M(0)^8}t \right)^{1/4}.
\end{equation}
The lifetime $t_{\rm LT}$ of a thin black ring with mass $M$ is
\begin{equation}
t_{\rm LT}\approx \left(\frac{27\pi\lambda}{4}\right)^2
\left(\frac{M}{M_p}\right)^2t_p,
\end{equation} 
where $M_p$ and $t_p$ are the Planck mass $(\hbar^2/G)^{1/3}$ 
and the Planck time $\hbar/M_pc^2$, respectively. 
The time scale is proportional to $M^2$, and 
this dependence on $M$ is same as that of the five-dimensional
Schwarzschild black hole. 
However, because of the prefactor $(27\pi\lambda/4)^2$, 
the lifetime of a thin black ring with $\lambda\lesssim 10^{-2}$ 
is much shorter than that of the five-dimensional Schwarzschild black hole 
with the same mass.

%
%======================================%
%<<<<<<<<<<<< SECTION V  >>>>>>>>>>>>>>%
%======================================%
%
%
%
%
\section{Energy and angular spectra}

In this section, we study 
the spectra of energy and angular momentum 
of radiated particles in the evaporation
of a thin Emparan-Reall black ring.

\subsection{Formulas for energy and angular spectra}

For a thin Emparan-Reall black ring with $a_\ast=0$,  
the emission rates of energy and angular momentum 
are given by Eqs.~\eqref{Energy-RadiationRate-without-prime}
and \eqref{AMpsi-RadiationRate-without-prime} with $dJ_{\phi}/dt=0$.
In Sec.~\ref{Sec:Simplification}, we simplified these formulas
by performing the Lorentz transformation from $(\tilde{\omega},\tilde{k})$ to 
$(\tilde{\omega}^\prime, \tilde{k}^\prime)$
through Eq.~\eqref{Lorentz tr}. Physically, this corresponds
to the transformation from the boosted frame to the unboosted frame. 
Therefore, when we speak about the spectra,
the two kinds of spectra can be considered:
The spectra in the boosted frame (with respect to $\omega$)
and those in the unboosted frame (with respect to $\omega^\prime$). 
In this paper, we prefer to analyze the spectra
in the {\it boosted} frame, because the spectra in the boosted
frame correspond to
those for a distant observer at rest in the original black ring
spacetime.  For this reason, we rewrite 
Eqs.~\eqref{Energy-RadiationRate-without-prime}
and \eqref{AMpsi-RadiationRate-without-prime}
in order to match them to our purpose here. 
Because the integrand does not depend on $m$ in the case of $a_\ast = 0$, 
we take summation with respect to $m$ in advance, 
\begin{equation}
\sum_{m=-\infty}^{\infty} g^{(m)}(\tilde{\omega},\tilde{k}) 
= 
\sum_{\ell=0}^{\infty}\left(2\ell+1\right) 
\Gamma_{\ell}(\tilde{\omega},\tilde{k}),
\end{equation}
and change the order of integration with respect to 
$\tilde{\omega}$ and $\tilde{k}$ as
\begin{equation}
\int_{-\infty}^\infty d\tilde{k} \int_{|\tilde{k}|}^{\infty}d\tilde{\omega}
= \int_{0}^{\infty}d\tilde{\omega} \int_{-\tilde{\omega}}^{\tilde{\omega}}d\tilde{k}.
\end{equation}
Then, the formulas for the emission rates become
\begin{equation}
\label{reformulated_dMdt}
-\frac{dM}{dt} = \frac{R}{\sqrt{2}\pi M_K^3}
\int_{0}^{\infty}d\tilde{\omega}
\sum_{\ell=0}^{\infty}\left(2\ell+1\right)
\int_{-\tilde{\omega}}^{\tilde{\omega}}d\tilde{k} 
\frac{\tilde{\omega}\Gamma_{\ell}(\tilde{\omega},\tilde{k})}{e^{(\tilde{\omega}-\tilde{k}V)/\tilde{T}}-1},
\end{equation}
\begin{equation}
\label{reformulated_dJpsidt}
-\frac{dJ_\psi}{dt} 
= \frac{R^2}{\pi M_K^3}
\int_{0}^{\infty}d\tilde{\omega}
\sum_{\ell=0}^{\infty}\left(2\ell+1\right)
\int_{-\tilde{\omega}}^{\tilde{\omega}}d\tilde{k} 
\frac{\tilde{k}\Gamma_{\ell}(\tilde{\omega},\tilde{k})}{e^{(\tilde{\omega}-\tilde{k}V)/\tilde{T}}-1}.
\end{equation}
Therefore, the energy and angular spectra are written as 
\begin{equation}
-\frac{d^2M}{dtd\tilde{\omega}}
= \frac{R}{\sqrt{2}\pi M_K^3}
\frac{dI_M}{d\tilde{\omega}}, 
\qquad 
-\frac{d^2J_\psi}{dtd\tilde{\omega}}
= \frac{R^2}{\pi M_K^3}
\frac{dI_{J}}{d\tilde{\omega}},
\end{equation}
with the definitions
\begin{equation}
\frac{dI_M}{d\tilde{\omega}} 
:= \sum_{\ell=0}^{\infty}\left(2\ell+1\right)
\int_{-\tilde{\omega}}^{\tilde{\omega}}d\tilde{k} 
\frac{\tilde{\omega}\Gamma_{\ell}(\tilde{\omega},\tilde{k})}{e^{(\tilde{\omega}-\tilde{k}V)/\tilde{T}}-1},
\label{Def:Rescaled-energy-spectrum}
\end{equation}
\begin{equation}
\frac{dI_{J}}{d\tilde{\omega}} 
:= \sum_{\ell=0}^{\infty}\left(2\ell+1\right)
\int_{-\tilde{\omega}}^{\tilde{\omega}}d\tilde{k} 
\frac{\tilde{k}\Gamma_{\ell}(\tilde{\omega},\tilde{k})}{e^{(\tilde{\omega}-\tilde{k}V)/\tilde{T}}-1}.
\label{Def:Rescaled-angular-spectrum}
\end{equation}
The quantities $dI_M/d\tilde{\omega}$ and $dI_{J}/d\tilde{\omega}$
are interpreted as the rescaled energy and angular spectra.  

We also would like to compare the energy spectrum of a thin black ring
with that of a four-dimensional Schwarzschild black hole.
The energy spectrum of evaporation 
of a four-dimensional Schwarzschild black hole
with a mass $M_S=M_K/G_4$, where $G_4$ is the four-dimensional gravitational constant, is given by
\begin{equation}
\label{ES_Sch}
-\frac{d^2M_S}{dtd\tilde{\omega}}
=\frac{1}{M_K^2}
\frac{dI^{\rm (BH)}_M}{d\tilde{\omega}} \quad \textrm{with} \quad
\frac{dI^{\rm (BH)}_M}{d\tilde{\omega}} 
:= \frac{1}{2\pi}\sum_{\ell=0}^{\infty}\left(2\ell+1\right)
\frac{\tilde{\omega}\Gamma^{\rm (BH)}_{\ell}(\tilde{\omega})}{e^{\tilde{\omega}/\tilde{T}^\prime}-1},
\end{equation}
where $\Gamma^{\rm (BH)}_{\ell}(\tilde{\omega})$ is the greybody factor 
for a massless scalar field in a four-dimensional Schwarzschild spacetime.
Here, $dI_{M}^{\rm (BH)}/d\tilde{\omega}$ is the rescaled energy spectrum. 
The trivial difference between the two energy emission rates
\eqref{reformulated_dMdt} and \eqref{ES_Sch} is that the
black ring evaporation is different by a factor of 
$\sim R/M_K\sim 1/\lambda\gg 1$
compared to the four-dimensional black hole evaporation.
This is because a large number of the Kaluza-Klein modes
contribute to the black ring evaporation,
while only massless modes contribute to the evaporation of a
four-dimensional Schwarzschild black hole. In the following,
we discuss the difference between the rescaled energy spectra
$dI_M/d\tilde{\omega}$ and $dI_M^{\rm (BH)}/d\tilde{\omega}$
apart from this trivial difference of $O(R/M_K)$.

\subsection{Numerical results}

Now, we present the numerical results. 

\subsubsection{Energy spectrum}

%===========<Figure3>============%
%
\begin{figure}[tb]
\begin{tabular}{cc}
\includegraphics[width=0.6\textwidth]{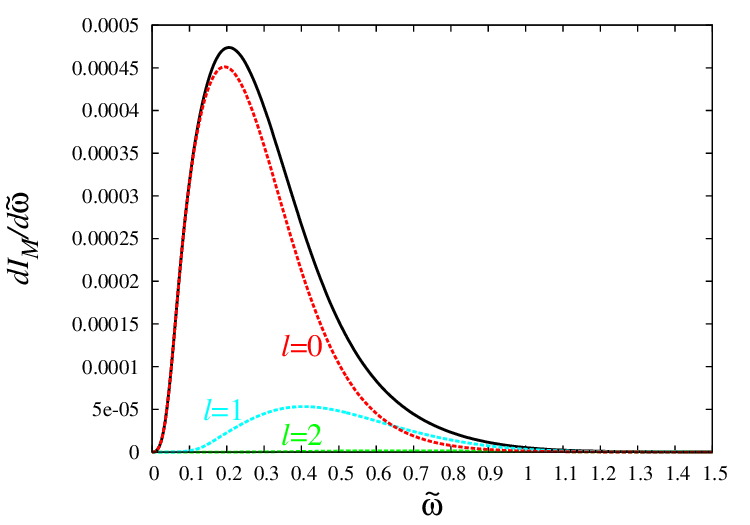}
\end{tabular}
\caption{
The rescaled energy spectrum $dI_M/d\tilde{\omega}$ as a function of $\tilde{\omega}$
together with the contributions of different quantum numbers $\ell=0,1$, and $2$.
This profile is proportional to the energy spectrum. 
}
\label{Fig:Energy_Spectrum}
\end{figure}
%
%=================================%

Figure~\ref{Fig:Energy_Spectrum} shows the rescaled energy spectrum 
$dI_M/d\tilde{\omega}$ 
of emitted particles from a thin Emparan-Reall black ring
as a function of $\tilde{\omega}$.
The contributions from different quantum numbers $\ell=0$, $1$, and $2$
are also plotted. Only the $\ell = 0$ and $1$ modes give the dominant
contributions for the energy spectrum. 
The data for the higher multipole modes $\ell\ge 3$
are not plotted because they are tiny and invisible.

%===========<Figure4>============%
%
\begin{figure}[tb]
\begin{tabular}{cc}
\includegraphics[width=0.6\textwidth]{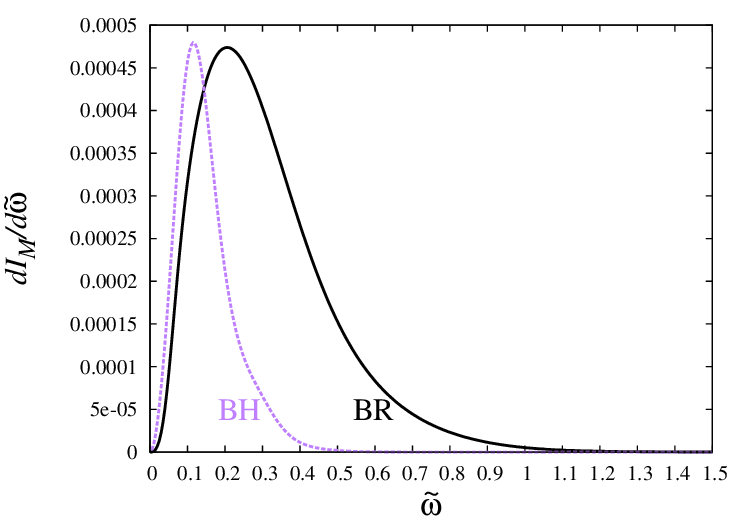}
\end{tabular}
\caption{
The rescaled energy spectra $dI_M/d\tilde{\omega}$ for a thin black ring 
and $dI^{\rm (BH)}_M/d\tilde{\omega}$ 
for a four-dimensional Schwarzschild black hole 
as a function of $\tilde{\omega}$. 
}
\label{Fig:Energy_Spectrum_compare}
\end{figure}
%
%=================================%

Let us discuss the feature of the energy spectrum of the black ring evaporation
by comparing it with that of the evaporation of a four-dimensional
Schwarzschild black hole. 
The numerical data of the 
energy spectra for a black ring and for a four-dimensional black hole
are plotted in Fig.~\ref{Fig:Energy_Spectrum_compare}.
First, we focus attention to the
low-frequency region, $\tilde{\omega}\ll 1$. 
In this region, the energy spectrum for the black ring
evaporation grows more slowly compared to that for the four-dimensional
black hole as $\tilde{\omega}$ is increased.
This feature can be explained by the approximate analysis as follows.  
In this region, the energy spectrum is approximately 
determined only by the $\ell = 0$ mode. 
Since the field equation in the unboosted frame is
identical to the Klein-Gordon equation with mass $k^\prime$,
we can use Unruh's approximate formula \cite{Unruh:1976fm} 
for the greybody factor,
\begin{eqnarray}
\Gamma_0 
&\approx& 
\frac{32\pi(1+v^{\prime 2})\tilde{\omega}^{\prime 3}}
{1-\exp[-2\pi\tilde{\omega}^\prime(1+v^{\prime 2})/v^\prime]}
\nonumber\\
&\approx&
16\tilde{\omega}^\prime\sqrt{\tilde{\omega}^{\prime 2}-\tilde{k}^{\prime 2}}
+16\pi\tilde{\omega}^\prime
\left(2\tilde{\omega}^{\prime 2}-\tilde{k}^{\prime 2}\right)+\cdots,
\label{Approximate-Greybody-Unruh}
\end{eqnarray}
for the $\ell=0$ mode, where 
$v^\prime:=\sqrt{1-\tilde{k}^{\prime 2}/\tilde{\omega}^{\prime 2}}$ 
is the velocity at infinity. Transforming this formula into the boosted
frame and substituting it into Eq.~\eqref{Def:Rescaled-energy-spectrum},
we find 
\begin{equation}
\frac{dI_M}{d\tilde{\omega}} \approx \tilde{\omega}^3.
\end{equation}
On the other hand, the approximate behavior of 
${dI_M^{\rm (BH)}}/{d\tilde{\omega}}$ for $\tilde{\omega}\ll 1$ 
for the four-dimensional
Schwarzschild black hole is derived as
\begin{equation}
\frac{dI_M^{\rm (BH)}}{d\tilde{\omega}} \approx \pi^{-2}\tilde{\omega}^2.
\end{equation}
This explains the slower growth
of the rescaled energy spectrum for the black ring 
compared to that for the four-dimensional black hole.

Next, we discuss the behavior in the high-frequency region
$\tilde{\omega}\gg 1$. In this case, the greybody
factor for a sufficiently small $\ell$ is approximately unity (see 
Fig.~\ref{Fig:gf} and also 
Eq.~\eqref{Eq:greybody} in Appendix~A), and therefore, 
the contribution from a mode with a sufficiently small $\ell$ 
is approximated as
\begin{equation}
\int_{-\tilde{\omega}}^{\tilde{\omega}} 
d\tilde{k}\frac{\tilde{\omega}\Gamma_\ell(\tilde{\omega},\tilde{k})}
{e^{(\tilde{\omega}-\tilde{k}V)/\tilde{T}}-1}
\approx 
\frac{\tilde{\omega}}{8\pi}{e^{-8(\sqrt{2}-1)\pi\tilde{\omega}}}.
\label{HighFrequency-OneMode}
\end{equation}
Since the number of the modes that
contribute to the energy spectrum is $O(\tilde{\omega}^2)$,
we have
\begin{equation}
\frac{dI_M}{d\tilde{\omega}}
\sim
{\tilde{\omega}^3}{e^{-8(\sqrt{2}-1)\pi\tilde{\omega}}}
\label{EnergySpectrum-HighFrequency-BlackRing}
\end{equation}
for $\tilde{\omega}\gg 1$ as an order estimate. 
On the other hand, for a four-dimensional
Schwarzschild black hole, we have
\begin{equation}
\frac{dI_M^{\rm (BH)}}{d\tilde{\omega}}
\sim
{\tilde{\omega}^2}{e^{-8\pi\tilde{\omega}}}.
\label{EnergySpectrum-HighFrequency-BlackHole}
\end{equation}
The remarkable difference of the black ring
formula~\eqref{EnergySpectrum-HighFrequency-BlackRing} 
from the black hole formula~\eqref{EnergySpectrum-HighFrequency-BlackHole}
is the presence of the factor $\sqrt{2}-1\approx 0.414$ 
in the argument of the exponential function. Because of this factor,
the energy spectrum for the black ring evaporation decays much more slowly
than that for the black hole as $\tilde{\omega}$ is increased.
We can also confirm this slower decay from 
our numerical data as shown in Fig.~\ref{Fig:Energy_Spectrum_compare}.
The origin of this factor is the argument 
$(\tilde{\omega}-\tilde{k}V)/\tilde{T}$ 
in the exponential function of the denominator 
in the left-hand side of Eq.~\eqref{HighFrequency-OneMode}.
In this formula, the momentum $\tilde{k}$ in the $z$ direction of the
boosted black string spacetime enters
like a chemical potential, and this ``chemical potential
term'' enhances the emission rate of 
particles with positive
momenta $\tilde{k}>0$. 
From the viewpoint of the original black ring spacetime, more number
of particles with positive angular momenta are emitted. 
Note that similar phenomena are observed in the evaporation
of Kerr and Myers-Perry black holes 
\cite{Kanti:2010mk,Ida:2005,Harris:2005,Duffy:2005,Casals:2008}: 
The energy emission rate
of a rotating black hole is also enhanced in the high-frequency regime
compared to that of a Schwarzschild(-Tangherlini) black hole
because of the effect of the chemical potential term.

The location of the peak has to be evaluated numerically. 
Our numerical result shows that 
the peak position is $\tilde{\omega} \simeq 0.21$ 
with the peak value $dI_M/d\tilde{\omega} = 4.73 \times 10^{-4}$. 
On the other hand, the peak position for 
the energy spectrum $dI_M/d\tilde{\omega}$ for a four-dimensional
Schwarzschild black hole is $\tilde{\omega} \simeq 0.12$. 
The peak of $dI_M/d\tilde{\omega}$ is located at a higher frequency 
(in the unit of $M_K$)
compared to that of $dI^{\rm (BH)}_M/d\tilde{\omega}$.
The difference in the peak positions comes from both 
the contribution from the Kaluza-Klein modes and 
the effect of the chemical potential term. 

To summarize, the energy spectrum of emitted particles from a black ring
shifts towards higher frequency domain compared to that from a four-dimensional
black hole with the same value of $M_K$.

\subsubsection{Angular spectrum}

%===========<Figure5>============%
%
\begin{figure}[tb]
\begin{tabular}{cc}
\includegraphics[width=0.6\textwidth]{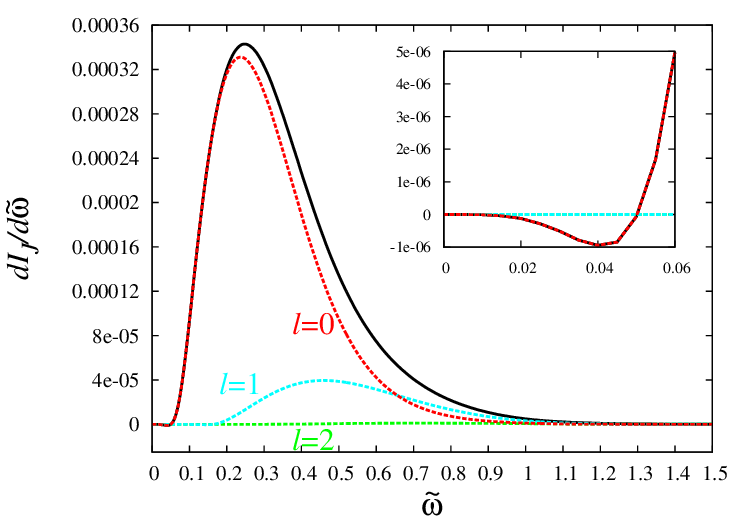}
\end{tabular}
\caption{
The rescaled angular spectrum $dI_{J}/d\tilde{\omega}$ as a function of $\tilde{\omega}$
together with the contributions of different quantum numbers $\ell=0,1$, and $2$.
The inset highlights the region $0 \le \tilde{\omega} \le 0.06$. 
This profile is proportional to the angular spectrum.}
\label{Fig:Angular_Spectrum}
\end{figure}
%
%=================================%

Now, we turn our attention to the angular spectrum. 
Figure~\ref{Fig:Angular_Spectrum} shows the rescaled angular spectrum 
$dI_{J}/d\tilde{\omega}$ as a function of $\tilde{\omega}$ 
together with the contributions of different quantum numbers 
$\ell=0,1$, and $2$. 
The modes $\ell\ge 3$ are not plotted for the same reason 
as the rescaled energy spectrum. 
Again, the $\ell = 0$ and $1$ modes give the dominant contributions
to the angular spectrum.

First, we discuss the behavior in the low-frequency region.
In this region, the spectrum is approximately determined only by 
the $\ell=0$ mode.
As we can see in the inset of Fig.~\ref{Fig:Angular_Spectrum}, 
the rescaled angular spectrum is negative 
for $\tilde{\omega}\lesssim 0.05$. 
This behavior can be confirmed also from the approximate analysis.
Substituting Unruh's approximate formula \eqref{Approximate-Greybody-Unruh}
for the greybody factor of the $\ell =0$
mode into Eq.~\eqref{Def:Rescaled-angular-spectrum},
we have
\begin{equation}
\frac{dI_J}{d\tilde{\omega}} 
\approx 
\left(\pi-\frac{8\sqrt{2}}{3}\right)
\tilde{\omega}^4
\approx 
-0.630\times\tilde{\omega}^4
\label{ApproximateAngularSpectrum-LowFrequency}
\end{equation}
after some calculation. In discussing the reason for this negativity,
there are two important effects: The chemical potential term and
the greybody factor. As discussed above,
the chemical potential term enhances the emission rate of particles
with positive $\tilde{k}$, and hence, tends to make the angular spectrum
positive. On the other hand, for a fixed Lorentz invariant
$\tilde{\omega}^2-\tilde{k}^2$, 
the greybody
factor is approximately proportional to the frequency
$\tilde{\omega}^\prime$ in the unboosted frame. 
Because $\tilde{\omega}^\prime = \sqrt{2}\tilde{\omega}-\tilde{k}$, 
the positive momentum $\tilde{k}$
decreases the transmission probability to infinity
for a given $\tilde{\omega}$. In other words, the relation 
$\Gamma(\tilde{\omega}, |\tilde{k}|)<\Gamma(\tilde{\omega}, -|\tilde{k}|)$
holds. The greybody factor suppresses
the emission of particles with positive momenta $\tilde{k}$,
and tends to make the angular spectrum negative.
Therefore, the two effects compete with each other.
At the leading order, the two effects cancel out and
there is no $O(\tilde{\omega}^3)$ term in 
Eq.~\eqref{ApproximateAngularSpectrum-LowFrequency}.
At the subleading order, 
the effect of the greybody factor is stronger than
the effect of the chemical potential, and this leads to the 
negative result of $O(\tilde{\omega}^4)$ in 
Eq.~\eqref{ApproximateAngularSpectrum-LowFrequency}.

Next, we discuss the behavior in the high-frequency region 
$\tilde{\omega}\gg 1$. 
As done in the discussion on the energy spectrum, we approximate the greybody
factor for a sufficiently small $\ell$ to be unity. Then,
the contribution from a mode with a sufficiently small $\ell$ 
is approximated as
\begin{equation}
\int_{-\tilde{\omega}}^{\tilde{\omega}} 
d\tilde{k}\frac{\tilde{k}\Gamma_\ell(\tilde{\omega},\tilde{k})}
{e^{(\tilde{\omega}-\tilde{k}V)/\tilde{T}}-1}
\approx 
\frac{\tilde{\omega}}{8\pi}{e^{-8(\sqrt{2}-1)\pi\tilde{\omega}}}.
\label{HighFrequency-AM-OneMode}
\end{equation}
Since the number of the modes that
contribute to the angular spectrum is $O(\tilde{\omega}^2)$,
an order estimate gives 
\begin{equation}
\frac{dI_J}{d\tilde{\omega}}
\sim
{\tilde{\omega}^3}{e^{-8(\sqrt{2}-1)\pi\tilde{\omega}}}.
\label{AngularSpectrum-HighFrequency-BlackRing}
\end{equation} 
This is the same behavior as that of the energy spectrum,
Eq.~\eqref{EnergySpectrum-HighFrequency-BlackRing}. The numerical
result also shows the similarity in the behavior of $dI_M/d\tilde{\omega}$
and $dI_J/d\tilde{\omega}$ in the high-frequency region.
Compare Figs.~\ref{Fig:Energy_Spectrum} and \ref{Fig:Angular_Spectrum}.

The peak position of the angular spectrum is numerically determined
to be $\tilde{\omega} \simeq 0.25$ with 
$dI_J/d\tilde{\omega}\simeq 3.43 \times 10^{-4}$.
This peak is located at a bit higher frequency than 
the peak location of the energy spectrum, and the peak values
of $dI_M/d\tilde{\omega}$ and $dI_J/d\tilde{\omega}$ have the same order.
To shortly summarize, a black ring emits positive angular momentum
except for a small region $\tilde{\omega}\lesssim 0.05$, and the
shape of the angular spectrum is similar to that of the energy spectrum.

%
%======================================%
%<<<<<<<<<<<< SECTION VI  >>>>>>>>>>>>>>%
%======================================%
%
\section{Summary}
\label{Sec:summary}

In this paper, we have studied the time evolution of evaporation
of a thin black ring under the assumption  
that only a massless scalar field is emitted in the Hawking radiation. 
In order to separate the Klein-Gordon equation in the background 
of a black ring metric, 
we have considered the thin-limit approximation, where 
the black ring metric is approximated by the boosted Kerr string metric. 
Then, we have given a set of equations, 
Eqs.~\eqref{time evolution of M}, \eqref{time evolution of Jpsi} 
and \eqref{time evolution of a}, 
that determines the quasistationary evaporation of a thin Pomeransky-Sen'kov black ring. 
In this setup, a black ring evaporates without
changing the thickness parameter $\lambda$. 
Also, we have analytically solved these equations 
in the case of an Emparan-Reall black ring 
and given its time evolution in Eqs.~\eqref{M_of_ER} and \eqref{Jpsi_of_ER},
with the factor $F(0)\simeq 0.239$
that has been determined by numerical calculation of the greybody factor.
In the evaporation, the shape of the Emparan-Reall black ring keeps 
similarity to its initial configuration. The lifetime of a black ring
is shorter by a factor of $O(\lambda^2)$ compared to a five-dimensional 
Schwarzschild black hole with the same initial mass.

We have also examined the energy and angular spectra of emitted
particles in the evaporation of a thin Emparan-Reall black ring. 
Compared to the energy spectrum for a four-dimensional 
Schwarzschild black hole,
the energy spectrum for a black ring shifts to high-frequency region.
In particular, 
the decay rate of the black ring spectrum is 
slower than that for a four-dimensional black hole
by a factor of $\sqrt{2}-1$ in the high-frequency region
because of the effect of the ``chemical potential'' term. 
It has also been shown that the angular spectrum 
has a similar shape to that of the energy spectrum except in the low-frequency
region $\tilde{\omega}\lesssim 0.05$ where the angular spectrum
becomes negative due to the effect of the greybody factors.

As a future work, we plan to study the time evolution of 
a Pomeransky-Sen'kov black ring with nonvanishing rotational
parameter $a_\ast$ along the $S^2$ direction. 
For this purpose, we have to calculate the functions
$F(a_*)$ and $H(a_*)$ of Eqs.~\eqref{Eq:def_Fa} and \eqref{Eq:def_Ha}, 
and therefore, developing the code for computing the greybody factor of 
the boosted Kerr string is required. 

%
%======================================%
%<<<<<<<<<<<< acknowledgments  >>>>>>>>>>>>>>%
%======================================%
%
\acknowledgments
 
This work was supported by the Grant-in-Aid for
Scientific Research (A) (22244030).

\appendix
%
%======================================%
%<<<<<<<<<<<< Appendix A  >>>>>>>>>>>>>>%
%======================================%
%
\section{DeWitt approximation}
\label{Appendix_A}

In order to check the validity of our numerical result~\eqref{Eq:result_F0}
of $F(0)$ in the case of the Emparan-Reall
black ring, we compute this value in
an approximate way. As the method of approximation, we adopt
the DeWitt approximation \cite{DeWitt:1975}
that was originally developed to
evaluate the contribution of the greybody factor
to the evaporation of a Schwarzschild black hole
(see p.~394 of Ref.~\cite{Frolov:1998} for a review).
In that study, the greybody factor was obtained
by appropriately reinterpreting the capture condition of null geodesics.
Although this approximation holds only for high-frequency
regime in a strict sense, it gives a rather
good result. In fact, the difference of the DeWitt approximation
from the numerical result is $\approx 6\%$. Compare the formula
of the mass-loss rate by the DeWitt approximation
(Eq.~(146) of Ref.~\cite{DeWitt:1975}) and the numerical value reported
in Ref.~\cite{Chambers:1997ai}.

In the spacetime of a five-dimensional
Schwarzschild string, a massless particle
with momentum in the $z$ direction effectively behaves
as a massive particle in a four-dimensional Schwarzschild
spacetime. Therefore, as the first step, we study
timelike geodesics in a four-dimensional Schwarzschild spacetime and
derive the capture condition. Then, we translate it to
the greybody factor for a massless scalar
field in a spacetime of a Schwarzschild string. Using this
result, we derive the value of $F(0)$ in the DeWitt approximation
by performing the summation and integration in
Eqs.~\eqref{Eq:def_g} and \eqref{Eq:F0_ER}.

The metric of a four-dimensional Schwarzschild spacetime is given by
\begin{equation}
ds^2=-f(r)dt^2+\frac{dr^2}{f(r)}+r^2(d\theta^2+\sin^2\theta d\phi^2),
\end{equation}
\begin{equation}
f(r)=1-\frac{2M_K}{r}.
\end{equation}
The geodesic motion of a massive particle
in the equatorial plane is governed by the following
equations:
\begin{equation}
f(r)\dot{t} = e,
\label{Eq:energy}
\end{equation}
\begin{equation}
r^2\dot{\phi}=j,
\label{Eq:AM}
\end{equation}
\begin{equation}
-f(r)\dot{t}^2 +
\frac{\dot{r}^2}{f(r)}+r^2\dot{\phi}^2
=
-1.
\label{Eq:fourth-eq}
\end{equation}
Here, $e$ and $j$ indicate the energy and angular momentum
per unit mass of the test particle,
and dot (\ $\dot{}$\ ) denotes the derivative
with respect to the particle's proper time $\tau$.
Substituting Eqs.~\eqref{Eq:energy} and \eqref{Eq:AM}
into Eq.~\eqref{Eq:fourth-eq}, we obtain
\begin{equation}
\dot{r}^2 + V(r) = e^2,
\end{equation}
where
\begin{equation}
V(r) = \left(\frac{j^2}{r^2}+1\right)\left(1-\frac{2M_K}{r}\right).
\end{equation}
Let us consider the situation where a test particle exists
in the neighborhood of the horizon and
moves toward outside (i.e. $\dot{r}>0$).
Denoting the peak value of
$V(r)$ as $V_{\rm peak}$,
the particle escapes to infinity if the condition
$V_{\rm peak} < e^2$ is satisfied. Conversely, a particle with
$V_{\rm peak} > e^2$ is reflected back to the black hole by the
centrifugal barrier.  After some calculation,
the condition $V_{\rm peak} = e^2$ is shown to be equivalent to
\begin{equation}
j = e\sqrt{F(e)}M_K,
\end{equation}
where
\begin{equation}
F(e) = \frac{27e^4 - 36 e^2 + 8 + e(9e^2-8)^{3/2}}{2e^2(e^2-1)}.
\end{equation}
Therefore, a particle escapes to infinity if $j < e\sqrt{F(e)}M_K$,
and it is reflected back to the black hole if $j > e\sqrt{F(e)}M_K$.
This condition is equivalent to the one obtained
in Ref.~\cite{Zakharov:1988}.
Here, $\sqrt{F(e)}$ varies from $4$ to $3\sqrt{3}$
as $e$ is increased from unity to infinity.

We use this result
in order to approximate the greybody factor
in the particle emission from the Schwarzschild string.
Here, we choose the unboosted frame, and as done
in Sec.~\ref{Sec:Simplification},
the quantities in this frame are indicated by prime $(\ {}^\prime\ )$.
Consider the emission of a quantum particle with
mass $k^\prime$,
angular frequency $\omega^\prime$, and angular quantum number $\ell$.
Replacing $j\to \ell/k^\prime$ and
$e\to \omega^\prime/k^\prime$
in the capture condition derived above, the particle
with $\ell \lesssim \omega^\prime\sqrt{F(\omega^\prime/k^\prime)}M_K$
escapes to infinity and that with
$\ell \gtrsim \omega^\prime\sqrt{F(\omega^\prime/k^\prime)}M_K$
falls back to the horizon.
Therefore, the greybody factor is approximated by
\begin{equation}
\Gamma^\prime_{\ell m}(\tilde{\omega}^\prime,\tilde{k}^\prime) \approx
\theta(\tilde{\omega}^\prime
\sqrt{F(\tilde{\omega}^\prime/\tilde{k}^\prime)}-\ell),
\label{Eq:greybody}
\end{equation}
where $\theta(u)$ denotes the Heaviside step function,
and we introduced $\tilde{\omega}^\prime = M_K \omega^\prime$
and $\tilde{k}^\prime = M_K k^\prime$ in the same manner as
Eq.~\eqref{Eq:def_tildeomega_etc}.
Note that in the massless particle limit,
$\tilde{k}^\prime/\tilde{\omega}^\prime\to 0$,
Eq.~\eqref{Eq:greybody} becomes $\Gamma_{\ell mn} \approx
\theta(3\sqrt{3}\tilde{\omega}^\prime-\ell)$, and
this agrees with the formula
in the original DeWitt approximation \cite{DeWitt:1975}.

Now, we evaluate the value of $F(0)$. The computation 
can be done with the formula given in 
Sec.~\ref{Sec:Evolution_of_ER_BR}. 
The quantity $g^\prime(\tilde{\omega}^\prime, \tilde{k}^\prime)$ 
in Eq.~\eqref{Eq:def_g} is
\begin{equation}
g^\prime(\tilde{\omega}^\prime, \tilde{k}^\prime)
=\sum_{\ell=0}^{\infty}
(2\ell +1)\theta(\tilde{\omega}^\prime
\sqrt{F(\tilde{\omega}^\prime/\tilde{k}^\prime)}- 
\ell)
\approx
\tilde{\omega}^{\prime 2}F(\tilde{\omega}^\prime/\tilde{k}^\prime).
\end{equation}
Here, the summation over $\ell$
was changed by integration as done by DeWitt.
Then, $F(0)$ can be calculated by substituting  this formula into 
Eq.~\eqref{Eq:F0_ER}. It is convenient to introduce
new variables $(x,y)$ by
$x=\tilde{k}^\prime/\tilde{\omega}^\prime$ and
$y=\tilde{\omega}^\prime$, and in these variables, 
analytic integration can be proceeded as
\begin{eqnarray}
I_1 &=& \int_0^\infty \frac{y^4dy}{e^{8\pi y}-1}\int_{-1}^{1}dxF(1/|x|)
\nonumber\\
&=&
\frac{\zeta(5)}{4096\pi^5}\left[88+33\sqrt{2}
\arcsin\left(\frac{2\sqrt{2}}{3}\right)-3\log 3\right].
\end{eqnarray}
Substituting this result into Eq.~\eqref{Eq:F0_ER},
we obtain
\begin{equation}
F(0)\approx 0.224.
\end{equation}
This value is fairly close to our numerical value in Eq.~\eqref{Eq:result_F0}:
Similarly to the original DeWitt approximation for the Schwarzschild  
black hole,
the approximate value is about 6\% smaller compared to the numerical  
value.
Therefore, the result of the DeWitt approximation supports the  
correctness
of our numerical calculation.

%---------   References   ---------%

%---------   References   ---------%

\end{document}